\pdfoutput=1

\documentclass[runningheads]{llncs}
\usepackage[numbers,sort&compress]{natbib}

\usepackage{times}
\usepackage{latexsym}

\usepackage[T1]{fontenc}

\usepackage[utf8]{inputenc}

\usepackage{microtype}

\usepackage{amsfonts}
\usepackage{amsmath}

\DeclareMathOperator*{\argmin}{arg\,min}

\usepackage{pifont}
\newcommand{\cmark}{\ding{51}}%
\newcommand{\xmark}{\ding{55}}%

\usepackage{graphicx}
\usepackage{booktabs}
\usepackage{multicol}
\usepackage{multirow}
\usepackage{todonotes}

\usepackage{listings}
\DeclareFixedFont{\ttb}{T1}{txtt}{bx}{n}{7.8pt} %
\DeclareFixedFont{\ttm}{T1}{txtt}{m}{n}{7.8pt}  %

\definecolor{deepblue}{rgb}{0,0,0.5}
\definecolor{deepred}{rgb}{0.6,0,0}
\definecolor{deepgreen}{rgb}{0,0.5,0}

\newcommand\pythonstyle{\lstset{
language=Python,
basicstyle=\ttm,
morekeywords={self},              %
keywordstyle=\ttb\color{deepblue},
emphstyle=\ttb\color{deepgreen},    %
stringstyle=\color{deepred},
frame=tb,                         %
showstringspaces=false
}}
\lstnewenvironment{code}[1][]
{
\pythonstyle
\lstset{#1}
}
{}

\title{Neural Approaches to Multilingual Information Retrieval}

 \author{Dawn Lawrie\inst{1}\orcidID{0000-0001-7347-7086} \and
 Eugene Yang\inst{1}\orcidID{0000-0002-0051-1535} \and \\
  Douglas W. Oard\inst{1,2}\orcidID{0000-0002-1696-0407} \and 
  James Mayfield\inst{1}\orcidID{0000-0003-3866-3013} 
}

 \authorrunning{D. Lawrie et al.}
\institute{
 HLTCOE. Johns Hopkins University, Baltimore  MD 21211, USA \\
 \email{\{eugene.yang,lawrie,mayfield\}@jhu.edu}
 \and
 University of Maryland, College Park MD 20742, USA  \email{oard@umd.edu}
 }

\begin{document}
\maketitle
\begin{abstract}
Providing access to information across languages
has been a goal of Information Retrieval (IR) for decades.
While progress has been made on Cross Language IR (CLIR)
where queries are expressed in one language and documents in another,
the multilingual (MLIR) task to create a single ranked list of documents across many languages is considerably more challenging. 
This paper investigates whether advances in neural document translation and pretrained multilingual neural language models enable 
improvements in the state of the art over earlier MLIR techniques.
The results show that although combining neural document translation with neural ranking yields the best Mean Average Precision (MAP), 98\% of that MAP score can be achieved with an 84\% reduction in indexing time by using a pretrained XLM-R multilingual language model to index documents in their native language, and that 2\% difference
in effectiveness is not statistically significant.  Key to achieving these results for MLIR is to fine-tune XLM-R using mixed-language batches
from neural translations of MS MARCO passages.

\end{abstract}

\section{Introduction}

With  advances in neural models for machine translation (MT) and Information Retrieval (IR), it is time to revisit
the problem of Multilingual IR (MLIR). Soon after Cross-Language IR (CLIR) was proposed as an information
retrieval task, research began on MLIR~\cite{oard1996survey}.
MLIR seeks to produce a total ordering over retrieved documents, regardless of language, such that the most useful documents appear at the top of the ranking. 
Assuming a searcher can consume multilingual information
(either directly or using MT),
the search engine should be able to return useful information regardless of the language of the document.

Much prior work on MLIR has involved 
subsetting documents by language, performing CLIR on each document set, and merging the results~\cite{peters2012multilingual}.
The advent of neural machine translation and 
neural IR using Multilingual Pretrained Language Models (MPLMs) creates new opportunities for
MLIR that we study here. 

If MT were perfect, translating all documents into the query language and searching monolingually might suffice. Indeed, our experiments confirm that for the high-resource languages with which we have experimented (English, French, German, Italian, and Spanish), using neural machine translation to convert each document into the query language is effective when used with neural ranking (in our experiments, ColBERT~\cite{khattab2020colbert}) fine-tuned on MS MARCO~\cite{bajaj2016ms}.
However, using neural MT in that way incurs substantial indexing costs because a GPU is required first to translate the document
and then again to encode it into dense vectors for neural IR.
Alternatively, we can use translations of MS MARCO to fine-tune an MPLM;
that approach is nearly as effective, not statistically different, and considerably faster at indexing time.
Our use of MS MARCO makes English a natural choice as the query language, but our approach is extensible to any query language for which suitable fine-tuning data exists.

This paper makes the following contributions:
    (1) Using a collection containing five high-resource European languages, we show that neural MT with neural IR achieves higher MAP and Precision at 10 scores than any other known MLIR technique, but that reliance on neural MT greatly increases the time required to index a collection.  
    (2) We show that extending the ColBERT-X~\cite{nair2022transfer} Translate-Train (TT) CLIR model to multiple languages achieves equivalent retrieval effectiveness with less than half the indexing time when used with mixed-language fine-tuning.
    (3) We show that some language bias in favor of query-language documents is present with all approaches, but that query-language bias is smaller with our Multilingual Translate-Train (MTT) implementation of ColBERT-X.

\section{Background}

We provide an overview of MLIR, followed by a brief review of traditional and neural IR.
The term ``multilingual'' has been used in several ways in IR. \citet{hull1996querying}, for example, note that it has been used to describe monolingual retrieval in multiple languages, as in~\citet{blloshmi-etal-2021-ir}, and it has also been used to describe CLIR tasks that are run separately in several languages~\cite{hc4,clef2001,clef2002,clef2003,ntcir2007overview}.
We adopt the Cross-Language Evaluation Forum (CLEF)'s meaning of MLIR: using a query to construct one ranked list in which each document is in one of several languages~\cite{peters2002importance}. We note that this definition excludes mixed-language queries and mixed-language documents, which are yet other cases to which ``multilingual'' has been applied.  

Five broad approaches to MLIR have been tried.  Among the earliest, \citet{rehder1997automatic} %
represented English, German and French documents in a learned trilingual embedding space, represented the query in the same embedding space, and then computed query-document similarity in the embedding space. %
The techniques and training data for creating multilingual embeddings were, however, too limited at the time to get good results from that technique.  More recently, \citet{SORG201226} %
garnered substantial attention by training embeddings on topically-related Wikipedia pages in English, German, French and Spanish. %
This paper extends this line of work.

A second approach by \citet{nie2002multilingual} %
indexed terms from all documents in their original language then created queries containing translations of the query terms in all target languages. %
With many document languages, this can lead to long queries.  A third approach
is to translate 
indexed terms into the query language at indexing time; the original queries can then be used directly to find similar (translated) content~\cite{magdy2011should, granell2014multilingual,rahimi2015multilingual}. %
We experiment with this approach as well.  This approach is, however, only practical when just
a few query languages are to be supported.  To address that limitation, the second and third approaches can be combined to create a fourth approach in which documents and query terms are each converted into one of a small number of indexing languages.
This has been called a ``pivot language'' approach, because in the limit all documents and queries can be translated into a single language.  

The fifth, and most widely studied, approach 
is to first use monolingual or bilingual retrieval
to create a ranked list for each document language,
and then to merge those ranked lists to construct a single result list~\cite{peters2012multilingual,si2008effective,tsai2008study}.
While this approach is architecturally similar to collection sharding, a widely-used approach to address efficiency, differences in collection statistics result in incompatible scores that require normalization prior to late fusion. Unfortunately, normalizing scores for collections across languages has been shown to be challenging~\cite{peters2012multilingual}.

Finally, one can simply show one ranked list per language to the user, as is done in the 2lingual search engine.\footnote{\url{https://www.2lingual.com/}}  This approach does not scale well beyond a small number of languages,
but it has the advantage of making it fairly clear to the searcher what the search engine has done.

Every MLIR ranked retrieval model
must rank the indexed documents given a query.  Traditional ranking methods such as computing inner products between the query and each indexed document containing a query term using sparse BM25~\cite{robertson2009probabilistic} term weights are fast, but neural IR methods yield better rankings~\cite{khattab2020colbert,karpukhin2020dpr,nair2022transfer} with more relevant documents earlier in the ranked list.  

This paper focuses on tradeoffs between effectiveness and efficiency.  Each technique described in this paper achieves ranking latency sufficient for interactive use (below 300 ms) on the collections that we experiment with,
but the time required to index the documents varies.
Indexing time consists of three components:
text processing (e.g., casing and tokenization),
machine translation,
and representation (e.g., \citet{mccarley1999should} and \citet{magdy2011should}).
Of these, neural MT is the slowest, so IR methods that do not require neural MT at indexing time have a substantial indexing time advantage (e.g., \citet{aljlayl2001effective}).
Our principal MLIR result is that MPLMs can achieve MAP close to the best results while producing substantial savings in indexing time.

We achieve this by extending the ColBERT-X~\cite{nair2022transfer} CLIR model to perform MLIR.  ColBERT-X combines three key ideas.  First, drawing insight from BERT~\cite{devlin2019bert}, it represents documents using contextual embeddings,
which better represent meaning than simple term occurrence. Second, using both multilinguality and improved pretraining from 
either multilingual BERT~\cite{xu-2021-global} or 
XLM-R~\cite{conneau2020xlmr}, ColBERT-X generates similar contextual embeddings for terms with similar meaning, regardless of language.  Third, drawing its structure from ColBERT~\cite{khattab2020colbert}, ColBERT-X limits ranking latency by separating query and document transformer networks,
allowing offline indexing.
ColBERT scores documents by focusing query term attention on the most similar contextual embedding in each document.  Our experiments confirm that this approach yields better MLIR MAP than does computation of inner products between classification tokens for the query and each document, an approach known as Dense Passage Retrieval (DPR)~\cite{karpukhin2020dpr}.

\section{Fine-Tuning MPLMs for MLIR}

Following \citet{nair2022transfer} we consider two
high-level approaches to  
fine-tuning for generalizing neural retrieval models to MLIR. Both approaches use existing MPLMs such as 
XLM-R~\cite{conneau2020xlmr} to encode queries and documents in multiple languages. We adapt the MPLM to MLIR via task-specific fine-tuning. 
These approaches are applicable to any retrieval model that is able to encode text using an MPLM. 

Consider a set of queries in a source language $\mathbf{L}_s$ and a set of documents in $m$ target languages $\mathbf{L}_t = \cup_{i=i}^m \mathbf{L}_i$.  We want to train a scoring function $\mathcal{M}_\Theta(q_{(s)}, d_{(t)}) \rightarrow 
\mathbb{R}$ for ranking documents with respect to a query. This paper denotes the language of an instance as a subscript $\bullet_{(l)}$.

\subsection{English Training (ET)}

Since MPLMs can encode text from many languages, we follow \citet{nair2022transfer} and only fine-tune the model monolingually. 
When processing queries, we transfer the model to MLIR zero-shot. Specifically, consider a loss function $\mathcal{L}$ (for example, cross-entropy),
\begin{equation*}
\Theta = \argmin_{\theta} \sum_{q, d} \mathcal{L}_\theta(q_{(s)}, d_{(s)}, r_{q,d})
\end{equation*}
where $q_{(s)}$ and $d_{(s)}$ are representations of the queries and documents and $r_{q,d}$ is the relevance judgment of document $d$ on query $q$, both in language $\mathbf{L}_s$, encoded by an MPLM. 
We use English as our query language because that is the language of MS MARCO.
We refer to this approach as ``English Training'' 
or ET.
However, this approach could equally well use any language for which similar extensive training data is available. 

Despite only exposing the model to text in $\mathbf{L}_s$ during fine-tuning,
the multilingual model can transfer its task model to other languages, as has been seen in prior CLIR work~\cite{nair2022transfer}.
However, such zero-shot language transfer is suboptimal because of (1) the lack of alignment objectives between languages during pretraining~\cite{yang2022c3}; and (2) differences in the representation of each language by the MPLM, which has been called \textit{the curse of multilinguality}~\cite{conneau2020xlmr,xu2021bibert}.
As we show in Section~\ref{sec:analysis:lang-bias}, such zero-shot transfer not only produces suboptimal retrieval effectiveness, it can also lead to language bias. 

\subsection{Multilingual Translate Training (MTT)}\label{sec:method:mtt}

To mitigate those issues, we propose a Multilingual Translate-Train (MTT) approach that generalizes the CLIR Translate-Train (TT) approach to MLIR~\cite{shi2019cross,nair2022transfer}.
To expose target languages $\mathbf{L}_1... \mathbf{L}_m$ to the model, we translate the monolingual training documents into each target language using MT. Specifically, the training objective can be expressed as
\begin{equation*}
\Theta = \argmin_{\theta} \sum_{q, d} \sum_{l=1}^{m} \mathcal{L}_\theta(q_{(s)}, d_{(l)}, r_{q,d})
\end{equation*}
This objective exposes the retrieval model to language pairs that it might see when processing queries, resulting in a more effective, better-balanced model.
We experiment with two batching approaches.
In Mixed-language (MTT-M), each batch contains documents in multiple languages, which encourages the model to learn similarity measures for all languages simultaneously.\footnote{Batches include the same query paired with document passages translated into each language.}
With Single-language (MTT-S), each batch contains only documents in one language, helping the model to learn retrieval for one language pair at a time. We found that MTT-M yields better retrieval effectiveness; thus, we present MTT-M as our main result. Section~\ref{sec:results:batch} compares the two approaches. 
In Section~\ref{sec:analysis:lang-bias}, we also demonstrate that MTT-M reduces language bias in MLIR. Implementation details can be found in Appendix~\ref{app:reproduce}

\begin{table*}[t]

\caption{Dataset statistics of CLEF 2001, 2002, and 2003. CLEF 2001 and 2002 share the document collection but have different queries. Numbers in parentheses are the number of topics in each query set. We report the number of documents judged relevant over all the topics in a particular year.}

\resizebox{\linewidth}{!}{
\centering
\begin{tabular}{c|rr|rr|rr|rr|rr|rr}
\toprule
Query & \multicolumn{2}{c|}{English} & \multicolumn{2}{c|}{German} 
         & \multicolumn{2}{c|}{Spanish} & \multicolumn{2}{c|}{French} 
         & \multicolumn{2}{c|}{Italian} & \multicolumn{2}{c}{Total}  \\
Set   & \# Rel. &  \# Docs 
	     & \# Rel. &  \# Docs 
	     & \# Rel. &  \# Docs 
	     & \# Rel. &  \# Docs 
	     & \# Rel. &  \# Docs 
	     & \# Rel. &  \# Docs 
	     \\
\midrule
2001 (50) 
&     856 &  \multirow{2}{*}{113,005}     %
&   2,130 &  \multirow{2}{*}{225,371}     %
&   2,694 &  \multirow{2}{*}{215,738}     %
&   1,212 &  \multirow{2}{*}{ 87,191}     %
&   1,246 &  \multirow{2}{*}{108,578}     %
&   8,138 &  \multirow{2}{*}{749,883} \\  %
2002 (50) 
&     821 &       %
&   1,938 &       %
&   2,854 &       %
&   1,383 &       %
&   1,072 &       %
&   8,068 &   \\  %
2003 (60) 
&   1,006 &  169,477     %
&   1,825 &  294,809     %
&   2,367 &  454,045     %
&     946 &  129,806     %
&      -- &       --     %
&   6,144 & 1,048,137 \\  %
\bottomrule
\end{tabular}
}

    \label{tab:data-stats}
\end{table*}

\section{Experiments}

One of the few test collections that currently supports MLIR evaluation with 
relevance judgments across multiple languages is from the the Cross-Language Evaluation Forum (CLEF). Following \citet{rahimi2015multilingual}
we use five document languages in the CLEF 2001-2002 collections~\cite{clef2001, clef2002} and four languages in the CLEF 2003 collection~\cite{clef2003}. 
Table~\ref{tab:data-stats} shows collection statistics.
We report performance for both title  and title+description queries,
also following \citet{rahimi2015multilingual}.
Because the number of query elements (subwords) is limited when encoding a query for dense retrieval,
we remove \textit{stop structure} to ensure that no query exceeds the length limit.
Stop structure includes phrases such as ``Find documents" and a limited stop-word list including ``on," ``the," and ``and."\footnote{For a complete list: \url{https://github.com/hltcoe/ColBERT-X/blob/main/scripts/stopstructure.txt}}

\subsection{Neural Retrieval Models}
We evaluate our proposed training approaches on two retrieval models -- ColBERT-X~\cite{nair2022transfer} and DPR-X~\cite{yang2022c3,zhang2021mrtydi},
which are multilingual variants of ColBERT~\cite{khattab2020colbert} and DPR~\cite{karpukhin2020dpr}. 
\citet{nair2022transfer} generalized the ColBERT~\cite{khattab2020colbert} model to CLIR, calling it ColBERT-X, by modifying the vocabulary space and replacing the monolingual pretrained language model with the MPLM XLM-RoBERTa~(XLM-R) Large (550M parameters)~\cite{conneau2020xlmr}. With proper training, ColBERT-X achieves state-of-the-art effectiveness in CLIR.
In this study, we integrate our proposed fine-tuning approaches with the ColBERT-X 
XLM-R %
implementation, which is based on the ColBERTv1 code base. 
We similarly adapted DPR~\cite{karpukhin2020dpr, yang2022c3},
a neural retrieval model that matches a single dense query vector
to a single dense document vector.
We name this model DPR-X. 
We use Tevatron~\cite{gao2022TevatronAE}, an open-source implementation of several neural end-to-end retrieval models in Python, for training, indexing, and retrieval. 

For training data, we use MS MARCO-v1~\cite{bajaj2016ms}, a commonly-used question-answering collection in English for fine-tuning
neural retrieval models.
For MTT, we use the publicly available mMARCO translations of MS MARCO~\cite{bonifacio2021mmarco},
fine-tuning using the ``small training triple'' (query, positive and negative document) file released by mMARCO's creators.
We trained all retrieval models with four GPUs (NVIDIA DGX and v100 with 32 GB Memory) with a per-GPU batch size of 32 triples for 200,000 update steps. All models are trained with half-precision floating points and optimized by the AdamW optimizer with a learning rate of $5\times 10^{-6}$. 

During indexing, documents are separated into overlapping spans of 180 tokens with a stride of 90~\cite{nair2022transfer}. We aggregate by MaxP~\cite{bendersky2008utilizing, dai2019deeper}, which takes the maximum score among the passages in a document as the document score. 

\subsection{Evaluation}\label{sec:exp:evaluation}
We report previously published results for
the state-of-the-art MULM~\cite{rahimi2015multilingual} system as a baseline for models that do not perform MT on the full collection. MULM is essentially an MLIR version of Probabilistic Structured Queries (PSQ)~\cite{darwish2003psq}.  PSQ  maps term frequency vectors from document to query language using a matrix of translation probabilities generated using statistical machine translation.
For MLIR, a translation matrix is created for each query-document language pair. 
The query likelihood model is used to score documents. Three key decisions led to good performance:
(1) estimating collection statistics based on translation probabilities; (2) estimating document length based on the translation and using that for smoothing; and (3) truncating the translation list at three.
As another baseline, we use BM25~($b=0.4$,  $k_1=0.9$) as implemented in Patapsco~\cite{costello2022patapasco} over neural machine translated documents (abbreviated ITD for Indexed Translated Documents).
For BM25, English queries and documents are tokenized by spaCy~\cite{spacy} and stemmed by the  NLTK~\cite{nltk} Porter stemmer (all supported by Patapsco).

For approaches that require document translation, we use directional MT models built on a transformer architecture (6-layer encoder/decoder) using Sockeye~2~\cite{sockeye2amta, sockeye2whitepaper}.
Measured by BLEU~\cite{papineni-etal-2002-bleu}, Sockeye~2 achieves state-of-the-art effectiveness
in each translation direction. Optimizations cut decoding time in half compared to Sockeye~1~\cite{sockeye1}.
We chose Sockeye~2 for its good trade-off between efficiency and effectiveness. 

To evaluate effectiveness on multiple languages in CLEF 2001-2002 and CLEF 2003, we combine the relevance judgments (qrels) for all languages for each query. In general, different languages have different numbers of relevant documents for each query. To evaluate models trained with English training data, we also translate the document sets into English with MT for indexing. 
Our main effectiveness measures are Mean Average Precision (MAP) and Precision at 10 (P@10). Both measures focus on the top of the rankings, and both were used by \citet{rahimi2015multilingual}, facilitating comparison between the neural approaches presented herein and prior state-of-the-art results. 

To evaluate language bias,
we count the number of relevant documents for a query across all languages,
and calculate recall at that level.
To compute the measure for a specific language,
we keep this level constant,
but ignore all documents in other languages (both in the MLIR results and in the relevance judgments).
We call the mean of this measure over all queries {\it Recall@MLIR-Relevant}.
When computing the mean, we omit from the calculation cases in which no relevant documents in that language are known
(recall is undefined in such cases).
This measure lies between 0 and 1, and values across that full range are achievable.   
We use the open source \texttt{ir-measures}~\cite{irmeasures}\footnote{\url{https://ir-measur.es/}}
package to compute all effectiveness measures.

\section{Results}

\begin{table*}[t]
\setlength\tabcolsep{0.55em}

\centering
\caption{Configurations of experiments identifying the pre-trained language model when applicable, 
the fine tuning data and process, the retrieval model, and the language of the indexed documents.
Under Fine-Tuning Data, MS MARCO refers to English MS MARCOv1, while mMARCO includes the translations into the various languages as well as the original English MS MARCOv1. A model that lists {\it{either}} under its 
Indexing Language can index either machine translated document (translation) or native documents in their various languages. 
} 
\label{tab:configs}
\resizebox{\linewidth}{!}{

\begin{tabular}{l|ccccc}
\toprule
       &   Language  &   Fine-Tuning  & Fine-Tuning  &   Retrieval  &    Indexing   \\
Name       &    Model &    Data &  Process &    Model &     Language  \\
\midrule
\footnotesize{MULM} & -- & -- & -- & PSQ & native \\
BM25-ITD & -- & -- & -- & BM25 & translation \\
ColBERT-X(ET) & XLM-R &  MS MARCO & ET & ColBERT-X & either \\
\footnotesize{ColBERT-X(MTT-M)} & XLM-R &  mMARCO & MTT-M & ColBERT-X & either \\
ColBERT-X(MTT-S) & XLM-R &  mMARCO & MTT-S & ColBERT-X & either \\
DPR-X(ET) & XLM-R &  MS MARCO & ET & DPR-X & either \\
DPR-X(MTT-M) & XLM-R &  mMARCO & MTT-M & DPR-X & either \\
ColBERT(ET) & BERT &  MS MARCO & ET & ColBERT & translation \\

\bottomrule
\end{tabular}
}

\end{table*}

We experiment with the Multilingual Translation Training (MTT) using two retrieval models
and compare them to two strong baseline retrieval models: BM25-ITD indexing translated documents and MULM indexing native documents;
these represent the state of the art on our test collections. 
Since per-query results for MULM have not been published
we perform significance tests only between our systems and the BM25+ITD baseline (the stronger of the two baselines).  
Table~\ref{tab:configs} summarizes the experiments that facilitate this analysis.
We first compare the effectiveness of our two batching strategies for MTT
before examining their effectiveness relative to the baselines.
Finally, we consider the trade-off between effectiveness and indexing time.

\subsection{Multilingual Batching for Fine-Tuning}\label{sec:results:batch}

\begin{table}[tb]
\setlength\tabcolsep{0.45em}

\renewcommand{\b}[1]{\textbf{#1}}
\renewcommand{\d}{$\dagger$}
\newcommand{\n}{\hphantom{$\dagger$}}
\centering

\caption{ColBERT-X MTT for Multiple or Single language training batches, indexing documents in their native language using title+description queries. \d~indicates significant improvement over MTT-S by paired $t$-test with 3-test Bonferroni correction ($p<0.05$).  
}

\begin{tabular}{l|rrr|rrr}
\toprule
      & \multicolumn{3}{c|}{\textbf{MAP}} 
      & \multicolumn{3}{c}{\textbf{P@10}} \\
      &  2001 & 2002 & 2003 &  2001 & 2002 & 2003 \\
\midrule
MTT-M &  
\b{0.462}\d &  \b{0.462}\d &  \b{0.461}\d &  
\b{0.704}   &  \b{0.752}   &  \b{0.653}   \\
\midrule
MTT-S &
0.422       &     0.405    &     0.433    &     
0.696       &     0.702    &     0.649    \\
\bottomrule
\end{tabular}

\vspace{-1em}

\label{tab:colbert-batch}
\end{table}

We compare two alternatives for fine-tuning the MTT condition and summarize the results with title+description queries in Table~\ref{tab:colbert-batch}. 
In 
all cases, mixed-language batches (MTT-M) produce more effective retrieval models than single-language (MTT-S). 
This is likely because, in MLIR, 
the model must rank documents from different languages together
instead of transferring trained models to other languages.
The outcome might be different if our goal were to perform CLIR over monolingual document collections.

\subsection{Effectiveness Relative to Baselines}

\begin{table*}[t]
\setlength\tabcolsep{0.25em}

\caption{MAP and P@10 on CLEF Title and Title+Description queries. Bold are best among a year; italics are best in a row ({\it{i.e.}}, with and without neural machine translation), $\dagger$ indicates significant difference from BM25+ITD by paired $t$-test with 16-test Bonferroni correction ($p<0.05$).
}

\centering
\renewcommand{\d}{$\dagger$}
\newcommand{\dd}{\hphantom{$\dagger$}}
\newcommand{\n}{\hphantom{$\dagger$}}
\renewcommand{\i}[1]{\textit{#1}}
\newcommand{\ib}[1]{\textbf{\textit{#1}}}

\resizebox{\textwidth}{!}{
\begin{tabular}{cc|rrrrrr||rrrrrr}
\toprule
    &     &  \multicolumn{6}{c||}{\textbf{MAP}} &  \multicolumn{6}{c}{\textbf{P@10}} \\
Query & 
      &  \multirow{2}{*}{MULM} &  \multirow{2}{*}{BM25} &  \multicolumn{2}{c}{ColBERT-X} & \multicolumn{2}{c||}{DPR-X} 
      &  \multirow{2}{*}{MULM} &  \multirow{2}{*}{BM25} &  \multicolumn{2}{c}{ColBERT-X} & \multicolumn{2}{c}{DPR-X} \\
Set & ITD  &        &        &     MTT-M\n&     ET\n&    MTT-M\n&     ET\n&        &        &    MTT-M\n&     ET\n&    MTT-M\n&     ET\n\\
\midrule
\multicolumn{14}{l}{Title Queries} \\
\midrule
\multirow{2}{*}{2001}
& \cmark &     --&\ib{0.398}&   0.377\n&  0.391\n&  0.338\n&  0.344\n&     --\n&\i{0.648}\n&  0.612\n&  0.596\n&  0.548\n&  0.584\n\\
& \xmark & 0.349&     --&   \i{0.360}\n&  0.322\n&  0.327\n&  0.298\d&\ib{0.650}\n&     --\n&  0.600\n&  0.588\n&  0.592\n&  0.570\n\\
\midrule
\multirow{2}{*}{2002}
& \cmark &     --&  0.337&   0.367\n&\ib{0.389}\n&  0.287\n&  0.304\n&     --\n&  0.618\n&  0.606\n&\ib{0.670}\n&  0.530\n&  0.596\n\\
& \xmark &  0.276&     --& \i{0.352}\dd&  0.333\n&  0.282\n&  0.277\n&   0.592\n&     --\n&  0.614\dd&\i{0.622}\n&  0.544\n&  0.556\n\\
\midrule
\multirow{2}{*}{2003}
& \cmark &     --&\ib{0.349}&   0.337\n&\ib{0.349}\n&  0.276\d&  0.266\d&     --\n&  0.595\n&  0.542\n&\ib{0.573}\n&  0.517\n&  0.497\d\\
& \xmark &  0.305&     --& \i{0.332}\dd&  0.290\n&  0.273\d&  0.247\d&      0.497\n&     --\n&\i{0.546}\n&  0.541\n&  0.527\n&  0.492\d\\
\midrule
\multirow{2}{*}{All}
& \cmark &     --&  0.361&   0.359\n&\ib{0.375}\n&  0.299\d&    0.302\d&     --\n&\ib{0.619}\n&   0.583\n&    0.611\n&  0.531\d&    0.554\d\\
& \xmark &  0.310&     --&\i{0.347}\n&    0.314\d&  0.293\d&    0.273\d&0.575\n&     --\n&   \i{0.584}\dd&    0.581\n&  0.553\d&    0.536\d\\

\midrule
\multicolumn{14}{l}{Title + Description Queries} \\
\midrule
\multirow{2}{*}{2001}
& \cmark &     --&  0.436&   0.472\n&\ib{0.477}\n&  0.365\n&  0.356\n&     --\n&  0.704\n&  0.718\n&\ib{0.754}\n&  0.658\n&  0.650\n\\
& \xmark &  0.387&     --&\i{0.462}\n&  0.405\n& 0.358\n& 0.324\d&      0.700\n&     --\n&  0.704\dd&\i{0.744}\n&  0.658\n&  0.644\n\\
\midrule
\multirow{2}{*}{2002}
& \cmark &     --&  0.398&   0.470\d&\ib{0.480}\d&  0.347\n&  0.332\n&     --\n&  0.696\n&\ib{0.774}\n&  0.770\n&  0.664\n&  0.620\n\\
& \xmark &  0.347&     --&\i{0.462}\n&  0.410\n&  0.335\n&  0.310\n&    0.666\n&     --\n&\i{0.752}\n&  0.720\n&  0.672\n&  0.640\n\\
\midrule
\multirow{2}{*}{2003}
& \cmark &     --&  0.394&\ib{0.419}\n&  0.410\n&  0.343\n&  0.328\d&     --\n&  0.615\n&  0.646\n&\ib{0.661}\n&  0.620\n&  0.600\n\\
& \xmark &  0.376&     --&\i{0.409}\n&  0.358\n&  0.338\n&  0.302\d&   0.563\n&     --\n&\i{0.653}\n&  0.637\n&  0.622\n&  0.575\n\\
\midrule
\multirow{2}{*}{All}
& \cmark &     --&  0.408&   0.451\d&\ib{0.453}\d&  0.351\d&    0.338\d&     --\n&  0.669\n&   0.709\n&\ib{0.725}\d&  0.646\n&    0.622\n\\
& \xmark &  0.368&     --&\i{0.442}\n&    0.390\n&  0.343\d&    0.312\d&  0.643\n&     --\n&\i{0.700}\dd&    0.697\n&  0.639\n&    0.617\n\\
\bottomrule
\end{tabular}
}
\label{tab:main-results}
\end{table*}

Our main effectiveness results are shown in Table~\ref{tab:main-results}. 
For ColBERT-X and DPR-X, MTT-M consistently improves effectiveness when retrieving documents in their native language
(i.e., {\it{without document MT}})
compared to English Training (ET).
Such improvements are seen in all three query sets,
and for both Title (T) and Title+Description (T+D) queries.
Differences are larger for MAP than P@10,
indicating that MTT-M affects more than just the top ranks. 

ColBERT-X MTT-M
numerically %
outperforms MULM for both query types and over all collections
in MAP and nearly all collections in P@10. 
With longer, more fluent title+description queries,
ColBERT-X MTT-M gives a larger improvement over MULM in both MAP and P@10,
indicating that XLM-R favors queries with more context.
Since DPR-X is less effective~\cite{yang2022c3},
MTT-M only brings its performance up to par with MULM.

With modern MT models,
we can improve MLIR effectiveness.
A common, yet strong, baseline of using BM25 to search over translated documents
yields substantial improvement over MULM in both MAP and P@10 with both query types. 
We argue that BM25+ITD is a proper baseline 
to
which future MLIR experiments should be compared.

\begin{table*}[t]
\setlength\tabcolsep{0.55em}

\centering
\caption{Monolingual ColBERT model using BERT-Large trained with ET and evaluated with translated documents. }
\label{tab:monolingual-results}

\resizebox{3.25in}{!}{
\begin{tabular}{l|rrrc||rrrc}
\toprule
              &  \multicolumn{4}{c||}{\textbf{MAP}} 
              &  \multicolumn{4}{c}{\textbf{P@10}} \\
Queries       &   2001 &   2002 &   2003 &    All &   2001 &   2002 &   2003 & All \\
\midrule
T &  0.397 &  0.367 &  0.362 &  0.375 &  0.592 &  0.646 &  0.583 &  0.606 \\
\midrule
T+D &  0.439 &  0.413 &  0.420 &  0.424 &  0.736 &  0.714 &  0.673 &  0.706 \\
\bottomrule
\end{tabular}
}
\vspace{-1em}

\end{table*}

We can also reduce neural IR to the monolingual case, training our retrieval model with English training and searching documents represented by English machine translations.
For both ColBERT and DPR,
an English-trained model (ET) indexing translated documents often yields better effectiveness than MTT-M indexing translated documents (ITD).
Furthermore, an English trained model indexing translated documents yields better effectiveness than MTT-M indexing documents in their native language;
however,
these differences are only statistically significant for CLEF 2002 on Title queries using a paired $t$-test with 3-test Bonferroni correction ($p < 0.05$).
We observe similar results with ColBERT using the BERT-Large pretrained LM trained under the same conditions except for using a learning rate of $3\times10^{-6}$ (the value suggested by the 
authors).
Compare Table~\ref{tab:monolingual-results} to ColBERT-X with English training, presented in Table~\ref{tab:main-results}.

\subsection{Preprocessing and Indexing Time}

\begin{table}[t]
\setlength\tabcolsep{0.45em}

\centering
\caption{ColBERT-X GPU hours for translating and indexing. BM25 does not use GPU.}

\begin{tabular}{lc|rrr|rrr}
\toprule
      &     &  \multicolumn{3}{c|}{CLEF2001-2002}  
               &  \multicolumn{3}{c}{CLEF2003} \\    
Model &  ITD & Translation & Index & Total &  Translation &  Index & Total \\
\midrule
BM25  
&  \cmark   &  55.0 &    -- &  55.0 &  68.6 &    --   &  68.6 \\
\midrule 
\multirow{2}{*}{ET}
&  \cmark   &  55.0 &   9.3 &  64.3 &  68.6 &  12.3   &  80.9 \\
&  \xmark   &    -- &   9.9 &   9.9 &    -- &  12.4   &  12.4 \\
\midrule
\multirow{2}{*}{MTT-S}
&  \cmark   &  55.0 &  16.9 &  71.9 &  68.6 &  19.0   &  87.6 \\
&  \xmark   &    -- &  16.7 &  16.7 &    -- &  21.9   &  21.9 \\
\midrule
\multirow{2}{*}{MTT-M}
&  \cmark   &  55.0 &  17.3 &  72.3 &  68.6 &  20.1   &  88.7 \\
&  \xmark   &    -- &  15.1 &  15.1 &    -- &  19.3   &  19.3 \\
\bottomrule
\end{tabular}

\label{tab:index_time}
\end{table}
\begin{figure*}[b]
    \centering
    \vspace{-1em}
    \includegraphics[width=\linewidth]{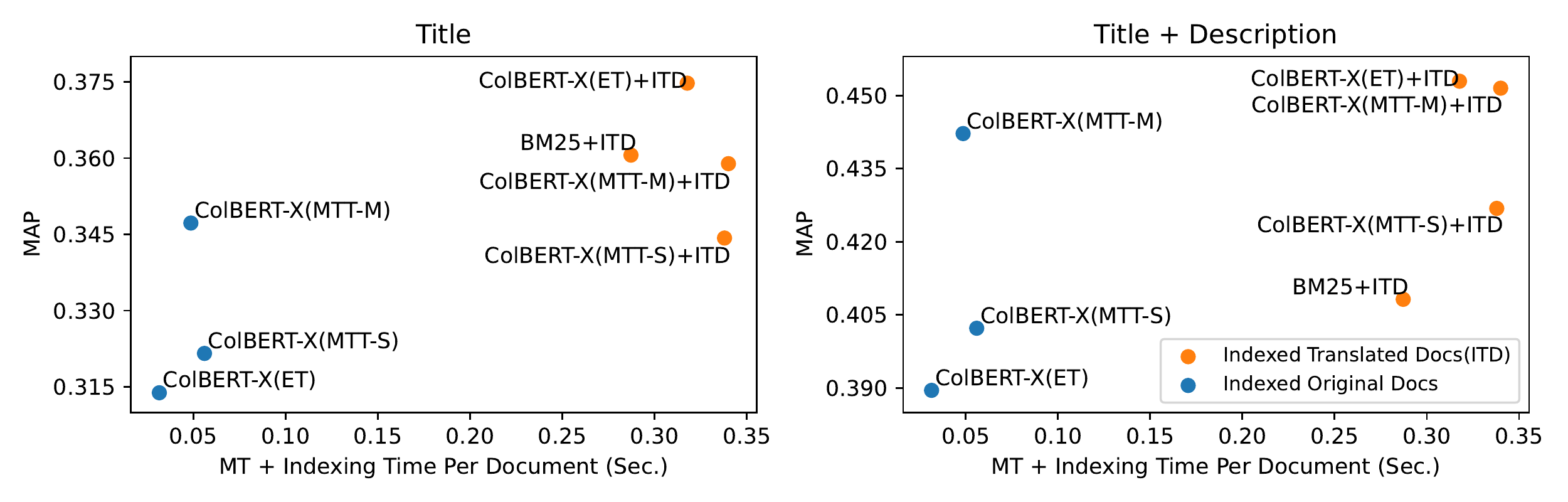}\vspace{-1em}
    \caption{Effectiveness (MAP) vs.~efficiency (per-document GPU indexing time in seconds) trade-off on CLEF 2001-2003. MAP scores (y-axis) for Title and Title+Description queries are disjoint ranges. The upper left is the optimal part of the chart. 
    }
    \label{fig:per-document-runtime-scatter}
\end{figure*}

Applying machine translation to entire document collections is expensive.
Table~\ref{tab:index_time} summarizes the cost for preprocessing and indexing the collection in GPU-hours for ColBERT-X and BM25.
We omit consideration of query latency here since all of our systems are sufficiently fast at query time for interactive use on collections of this size.
We refer the interested reader to \citet{santhanam2022plaid}.

This table reveals that differences in total indexing time between searching native and translated documents range from four to 6.5 times
depending on collection size and model.\footnote{Although Marian~\cite{junczys2018marian} is faster than Sockeye 2, benchmark results from Sockeye 1~\cite{sockeye1} and Sockeye 2~\cite{sockeye2whitepaper} confirm that Sockeye 2 is within a factor of 2 to 3 of Marian's speed, leaving our conclusions unchanged.} 
Despite that searching translated documents with monolingual retrieval models is more effective,
the computational cost of MT at indexing time is significantly higher;
one might choose not to bear this cost in exchange for the small and not statistically significant 
numerical gain in measured 
effectiveness 
over searching documents in their native language with MTT-M fine-tuning for title+description queries. 

We also see this trade-off on a per-document basis.
Figure~\ref{fig:per-document-runtime-scatter} shows that ColBERT-X with English training searching translated documents (\textit{ColBERT-X(ET)+ITD}) achieves the best effectiveness with both title (0.375 MAP) and title+description (0.453 MAP) queries.
However, it has a high preprocessing cost of 0.32 seconds per document,
whereas ColBERT-X trained with MTT-M searching documents in their native languages (\textit{ColBERT-X(MTT-M)})
requires under 0.05 seconds per document.
This is an 84\% reduction in preprocessing cost
at an apparent (but not statistically significant) cost of only 2\% in MAP with title+description queries.

\section{Analysis}

This section investigates our 
experimental results by breaking down the collection in two ways --
by document language, and by topic.

\subsection{Language Bias}\label{sec:analysis:lang-bias}

\begin{figure*}[b]
    \centering
    \vspace{-1.5em}
    \includegraphics[width=\linewidth]{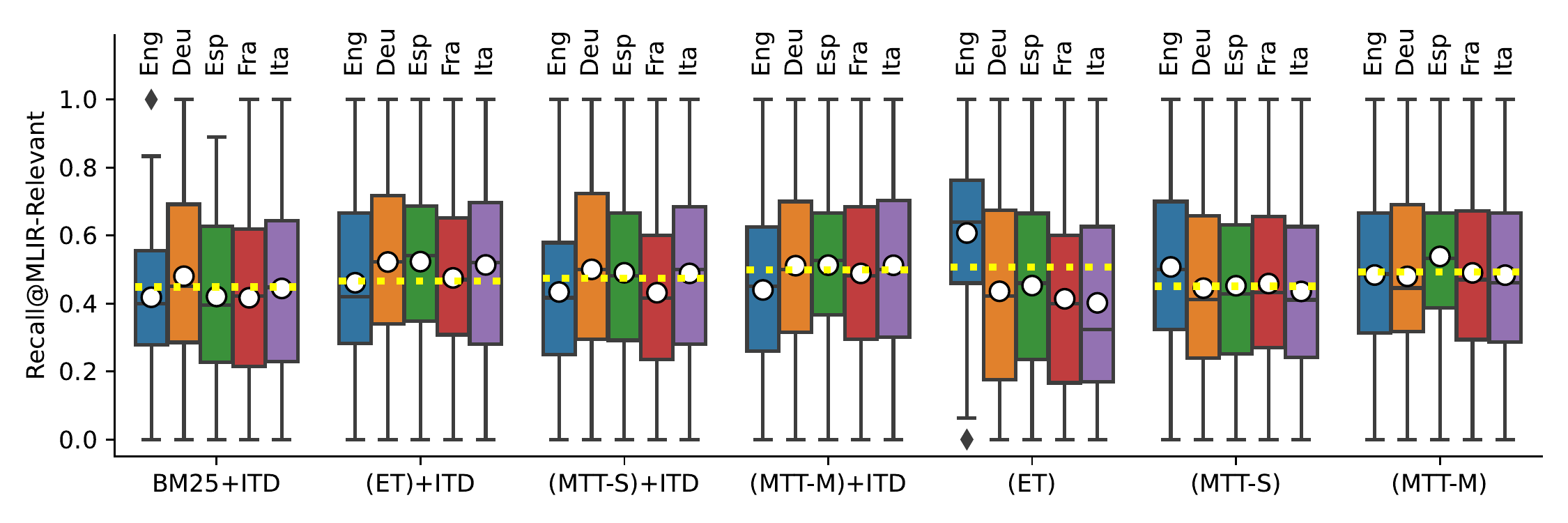}\vspace{-1em}
    \caption{R@MLIR-Relevant of BM25 and ColBERT-X variants for each language in CLEF2001-2002 with title+description queries. The yellow dashed line is the average over all languages, i.e., the R-Precision in MLIR. Outliers are defined as values beyond 1.5$\times$interquartile range. Horizontal black bars
    indicate the median and white circles indicate the mean. 
    }
    \label{fig:r-at-mlir-r}
    \vspace{-1em}
\end{figure*}

Since MPLMs are known to exhibit language biases~\cite{choudhury2021linguistically, kassner2021multilingual}, 
we investigate how retrieval models fine-tuned with our training schemes inherit or alleviate these biases. In MLIR we consider a model biased if it ranks a language's documents systematically higher or lower than those of another language. While MLIR is not a new
task, we are not aware of prior work that has examined language bias. Therefore we introduce two approaches to studying this phenomenon.
The first approach examines rates of relevant documents.
Since relevant documents are unevenly distributed across languages
(e.g., Spanish has more than three times as many known relevant documents as English among the CLEF 2001 topics,
averaging 54 vs.~17 relevant documents per topic, respectively), 
meaningful comparisons require us to focus on rates rather than on counts.
In this analysis, we focus on Recall@MLIR-Relevant (see Section~\ref{sec:exp:evaluation}), illustrating our analysis using the 100 title+description queries in CLEF 2001-2002 topics to characterize the coverage of relevant documents in each language (results on CLEF 2003 topics are similar).

Figure~\ref{fig:r-at-mlir-r} shows distributional statistics 
of Recall@MLIR-Relevant over topics 
by language and condition
that have at least one known relevant document in that language (96 for German, 97 for Spanish, 94 for Italian, 90 for French, 73 for English).  
When transferring a ColBERT-X model fine-tuned zero-shot with English training (i.e., ColBERT-X(ET)) to other languages, the model favors English documents due to the fine-tuning condition.
This results in a strong language bias in the retrieval results. Such biases can be ameliorated by fine-tuning with MTT. MTT-M appears to have more consistent behavior across languages compared to MTT-S, although the small apparent difference is not statistically significant. 
When indexing translated documents, Recall@MLIR-Relevant tends to be lower for English compared to other languages (though also not significantly). Since documents were translated sentence-by-sentence,
we hypothesize that indexing translated documents provides more synonym variety when decoding similar terms, resulting in document expansion;
this hypothesis requires more investigation, which we leave for future work. 

An alternative approach to investigating language bias is to assume that in a bias-free approach to MLIR, the scores for relevant documents would be drawn from the same underlying distribution. 
Using the 2-sample Kolmogorov-Smirnov test, the null hypothesis is that the two samples are drawn from the same distribution. For this analysis, we chose English as a reference and tested each topic with at least three relevant documents in each language. We then adjusted the p-values to account for multiple comparisons. We found that we could reject the null hypothesis for all languages and all configurations, indicating the document scores are not drawn from the same distribution based on language. Although some of this difference could result from differences in collection statistics
(i.e., with some languages better supporting the queries than others based on the numbers of relevant documents), 
the differences we observe across retrieval models indicate that there are retrieval model effects as well.
Notably, ColBERT-X(ET) retrieving documents in the native language has the largest percentage of topics with bias (from 15\% to 30\% depending on language pair), while all other configurations have no more than 12\% of topics exhibiting biased scores. 
This confirms the qualitative analysis above, which revealed that ColBERT-X(ET) over the documents in their native language had the most skewed rates of relevant documents. Future research will need to address language bias in document scores.

\subsection{Example Queries}

\begin{figure*}[tb]
    \centering
    \includegraphics[width=\linewidth]{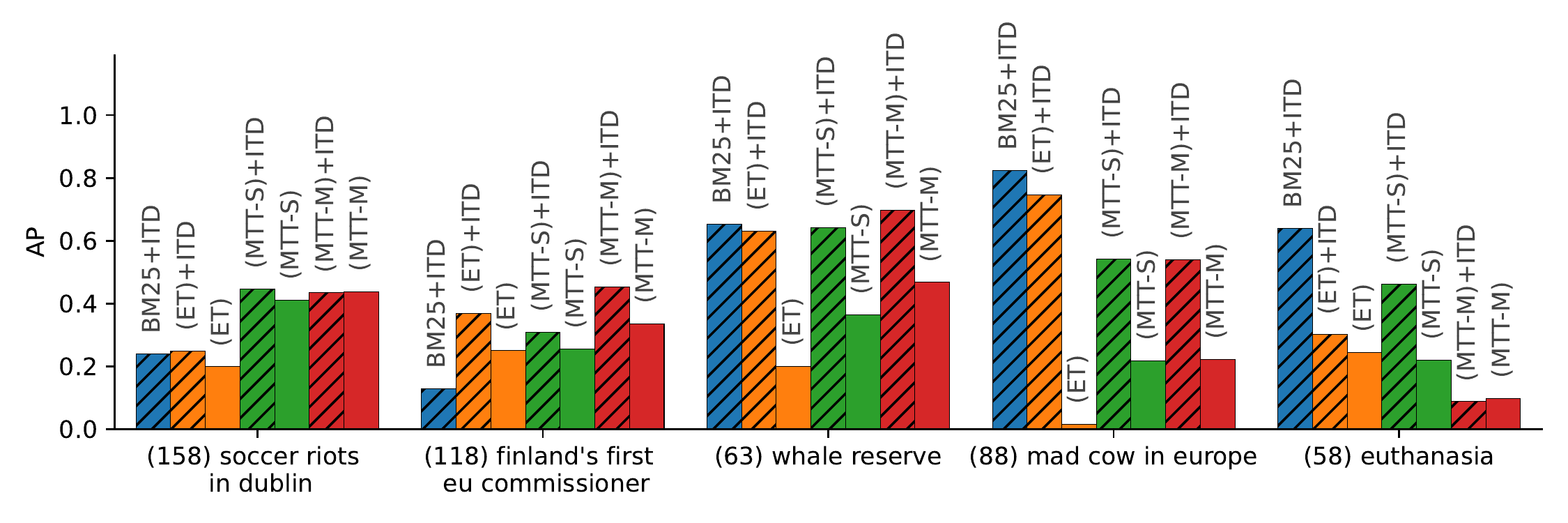}\vspace{-1.5em}
    \caption{Average Precision (AP) of BM25 and ColBERT-X on selected topics using title queries.}
    \label{fig:per_query}
    \vspace{-1em}
\end{figure*}

For more insight into differences among the algorithms, we show effectiveness on 
individual queries in Figure~\ref{fig:per_query}.
Our query selection here is not meant to be representative, but rather illustrative of phenomena that we see. 
For two topics on which ColBERT-X outperformed BM25 (topics 158 and 118),
the queries include terms that likely benefit from ColBERT-X soft term-matching -- ``soccer'' and ``commissioner'' respectively.
This term expansion effect has also been observed in monolingual retrieval with ColBERT. 

MT is particularly helpful for topics 63 and 88,
likely due to the quality of the translation for documents on these topics.
Especially for topic 88, English monolingual retrieval produces strong results.
Such behaviors indicate that the multilingual term matching in ColBERT-X is still not as effective 
on less common concepts like ``mad cow'' as is machine translation.

Topic 58 is an outlier. The term ``euthanasia'' is tokenized as a single token for BM25 but separated into \texttt{\_eu}, \texttt{thana}, and \texttt{sia} by the XLM-R tokenizer;
combined with the minimal context provided by a query,
this prevents ColBERT-X from matching properly across languages. 
Such diverse behaviors suggest room for further MLIR improvements using system combination. 
\section{Conclusion and Future Work}

This paper proposes the MTT training approach to MLIR that uses translated MS MARCO.
When searching non-English documents,
fine-tuning with MTT using mixed-language batches (MTT-M) enables neural models such as ColBERT and DPR to be more effective
than if fine-tuned on English MS MARCO. 
ColBERT-X with MTT-M is not statistically different from  monolingual English models applied to neural indexing-time translation of the collection into English, 
yet it achieves substantially better indexing time efficiency.
These results may not hold for more diverse sets of languages or when
MT is less effective; future work will examine the multilingual topics from the TREC 2022 the NeuCLIR track,\footnote{https://neuclir.github.io/} which judges the relevance of documents 
written in Chinese, Persian, and Russian.
Our observation that the retrieval method that yields the best retrieval effectiveness is query-dependent suggests future work on system combination, but our focus on efficiency and on language bias also calls attention to issues beyond retrieval effectiveness that will merit consideration in such a study.

\bibliography{anthology,custom}
\bibliographystyle{splncs04nat}

\appendix
\section{MTT Implementation Details}\label{app:reproduce}

As described in Section~\ref{sec:method:mtt}, MTT-M consists of examples with different languages in the training batches. We implement it by mixing the translated MS-MARCO triples round-robin. 
Specifically, each triple consists of an English query and positive and negative passages translated into the target languages. We constructed such triples using the translated documents provided by  mMARCO~\cite{bonifacio2021mmarco}. Each language results in a triple file of the same structure as \texttt{triples.train.small.tar.gz}.\footnote{\url{https://msmarco.blob.core.windows.net/msmarcoranking/triples.train.small.tar.gz}} 
The following Bash command creates a combined triple file that mixes all languages: 
\begin{code}
paste -d '\n' <(cat ./original_msmarco/triples.train.small.tsv)  \
              <(cat ./mmarco/french/triples.train.small.tsv)  \
              <(cat ./mmarco/german/triples.train.small.tsv)  \
              <(cat ./mmarco/italian/triples.train.small.tsv) \
              <(cat ./mmarco/spanish/triples.train.small.tsv) \
| cat > combined.tsv
\end{code}
Training with four GPUs and a per-GPU batch size of 32 triples guarantees that each batch consists of examples in different languages based on ColBERT-X's\footnote{\url{https://github.com/hltcoe/ColBERT-X/blob/main/xlmr\_colbert/training/lazy\_batcher.py}} batching scheme.

For MTT-S, we modified the ColBERT-X batching mechanism to load multiple triple files and supply a batch of examples from only one source file whenever the training process requests 
one. After each request, we switch the source triple file to ensure all languages are presented equally to the model during training.

\end{document}